\documentclass{moriond}

\def\be{\begin{equation}}
\def\ee{\end{equation}}
\def\bea{\begin{eqnarray}}
\def\eea{\end{eqnarray}}

\usepackage{amsmath}

\newcommand{\Photo}{}

\begin{document}
\vspace*{4cm}
\title{Precision Studies of the $\eta_c$ decay at BESIII}

\author{ Yijia Zeng on behalf of BESIII Collaboration  }

\address{Institute of High Energy Physics, Beijing 100049, People's Republic of China}

\maketitle\abstracts{
The lowest lying charmonium system $\eta_c$ has been observed for more than four decades. Studies of its production and decay properties provide an unique platform to investigate the inner structure of charmonium systems, hence improve our understanding of strong interaction in the charm sector. BESIII detector at the $e^+e^-$ BEPCII collider has already collected the world largest $J/\psi$ and $\psi(3686)$ data sets, based on which massive $\eta_c$ samples are produced via radiative transitions. This paper reviews recent precision studies of the $\eta_c$ decays at BESIII, including $J/\psi\to\gamma\eta_c$, $\eta_c\to\gamma\gamma$, and several $\eta_c$ hadronic decays.
}

\section{Introduction}


Charmonium indicates a bound state consists of a charmed quark and a charmed anti-quark. The first charmonium state $J/\psi$ was discovered at BNL~\cite{E598:1974sol} and SLAC~\cite{SLAC-SP-017:1974ind} in 1974. Since then, all the charmonium states below the open-charm threshold have been well-established~\cite{ParticleDataGroup:2024cfk}, and the measured spectrum is in good agreement with theoretical calculations based on the QCD-inspired potential models~\cite{Godfrey:1985xj}. Among them, the lowest lying system $\eta_c$ is of special importance. Compared to these complicated excited states like $\chi_{cJ}$ and $h_c$, $\eta_c$ is more friendly for the theoretical investigations. Massive approaches, including dispersion sum rules~\cite{Khodjamirian:1979fa}, QCD sum rules~\cite{Beilin:1984pf,Bodwin:2001pt}, relativistic quark models~\cite{Ebert:2002pp,Chen:2016bpj}, non-relativistic potential models~\cite{Deng:2016stx,Barnes:2005pb}, effective field theories~\cite{Brambilla:2005zw,Pineda:2013lta,Segovia:2021bjb,Brambilla:2018tyu,Yu:2019mce,Feng:2017hlu}, light-cone sum rules~\cite{Guo:2019xqa}, and lattice QCD~(LQCD)~\cite{Dudek:2006ej,Dudek:2009kk,Chen:2011kpa,Becirevic:2012dc,Donald:2012ga,Gui:2019dtm,Colquhoun:2023zbc,Meng:2024axn,CLQCD:2020njc,CLQCD:2016ugl,Liu:2020qfz,Dudek:2006ut,Meng:2021ecs}, have been employed in the $\eta_c$ relevant calculations. Several interesting observables, such as the transition rate of $J/\psi\to\gamma\eta_c$, two-photon decay width of $\eta_c\to\gamma\gamma$, and hyperfine mass spliting between $J/\psi$ and $\eta_c$, have been studied for several decades in both experimental and theoretical aspects, which provides an unique platform to investigate the inner structure of charmonium systems and improve our understanding of strong interaction in the charm sector.

\begin{figure}[htbp]
	\centering
	\setlength{\abovecaptionskip}{0.cm}
	\includegraphics[width=7cm]{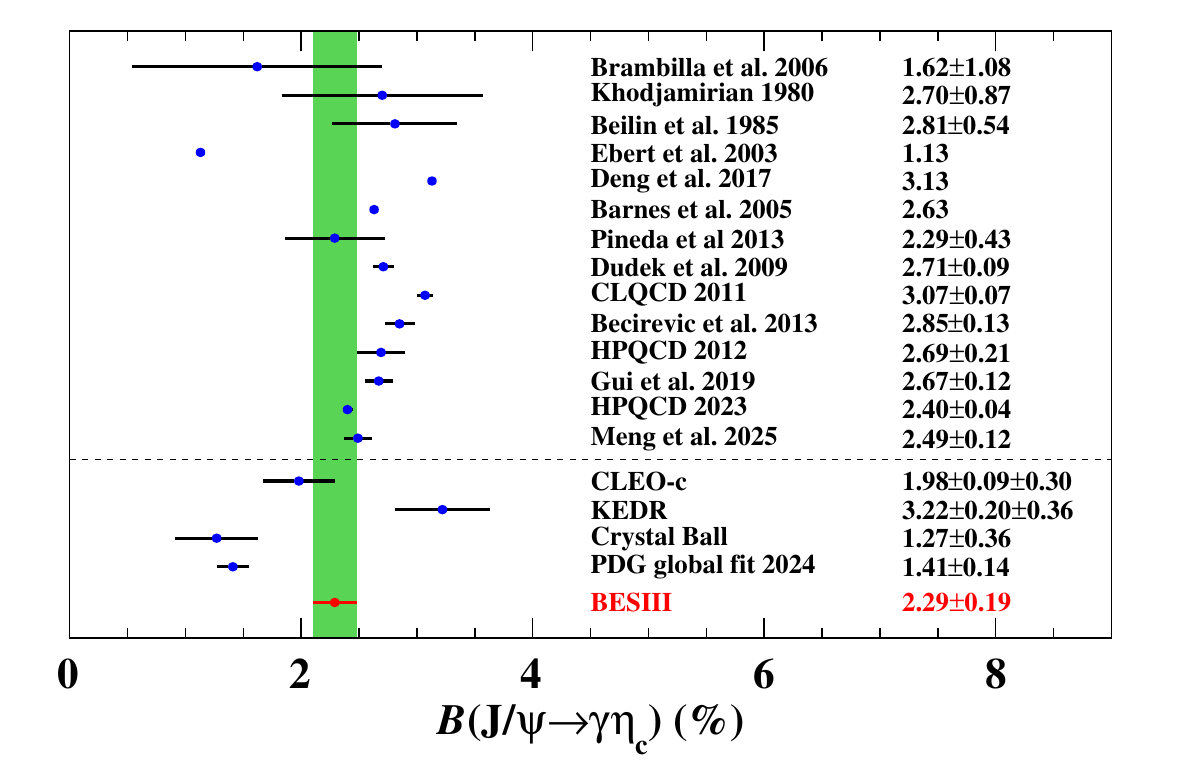}
	\includegraphics[width=7cm]{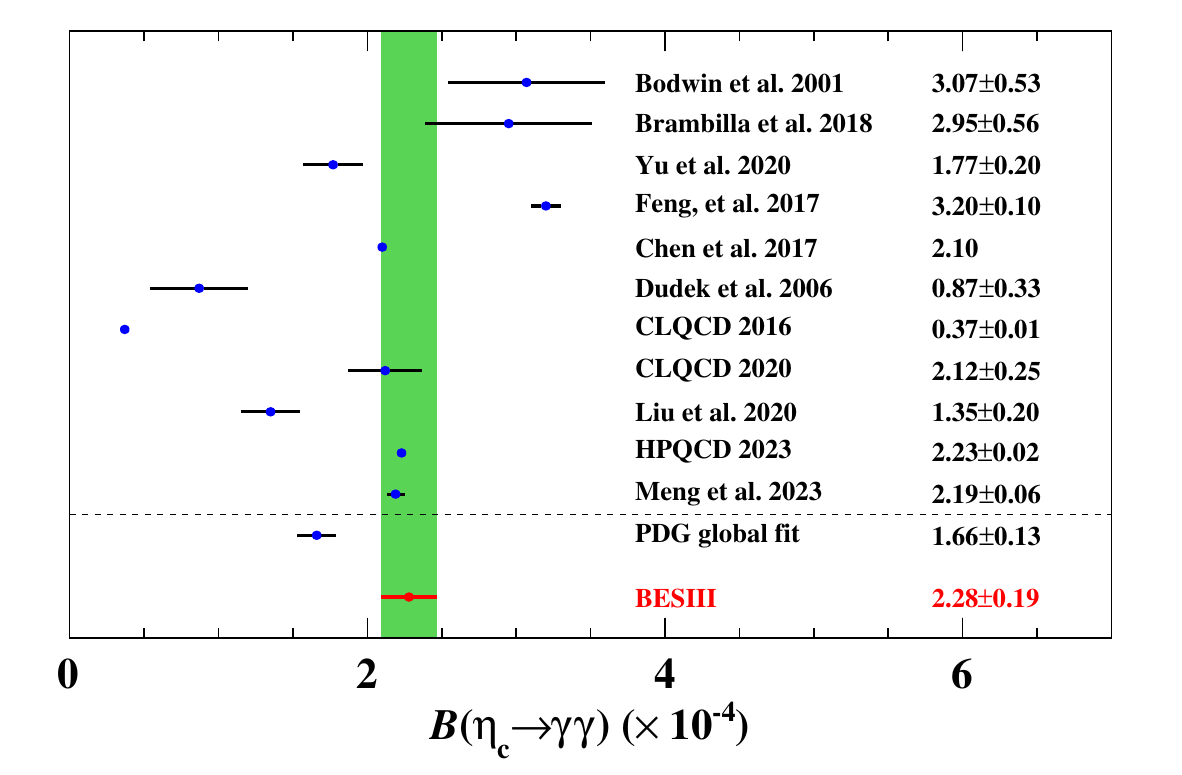}
	\caption{The comparison between experimental measurements and theoretical calculations on the branching fractions $\mathcal{B}(J/\psi\to\gamma\eta_c)$~(left) and $\mathcal{B}(\eta_c\to\gamma\gamma)$~(right).}
	\label{fig:compare_th_exp_bfpdg}
\end{figure}

However, there are still two puzzles for $\eta_c$. One is the large discrepancies between PDG global fit~\cite{ParticleDataGroup:2024cfk} and theories on the branching fractions $\mathcal{B}(J/\psi\to\gamma\eta_c)$ and $\mathcal{B}(\eta_c\to\gamma\gamma)$ shown in Fig.~\ref{fig:compare_th_exp_bfpdg}. One is that half of the decay modes of $\eta_c$ is unknown~\cite{ParticleDataGroup:2024cfk}. To solve these puzzles, more precision measurements on the $\eta_c$ decays are highly desired. Until now, BESIII has accumulated the world largest 10 Billion $J/\psi$~\cite{BESIII:2021cxx} and 2.7 Billion $\psi(3686)$~\cite{BESIII:2024lks} data sets, which allows rich $\eta_c$ productions via three radiative transitions $J/\psi\to\gamma\eta_c$, $\psi(3686)\to\gamma\eta_c$, and $\psi(3686)\to\pi^0 h_c, h_c \to \gamma \eta_c$, hence enables the precision studies of $\eta_c$. This paper reviews recent precision studies of the $\eta_c$ decays at BESIII, including $J/\psi\to\gamma\eta_c$, $\eta_c\to\gamma\gamma$, and several $\eta_c$ hadronic decays.

\section{Measurement of $\eta_c\to\gamma\gamma$ and $J/\psi\to\gamma\eta_c$}

Previously, the branching fraction $\mathcal{B}(\eta_c\to\gamma\gamma)$ is measured in two photon production $\gamma\gamma\to \eta_c$ and $p\bar{p}$ annihilation $p\bar{p}\to \eta_c \to \gamma\gamma$. Given the large discrepancy between experiments and theories, other measurement approaches for $\eta_c\to\gamma\gamma$ are desired.

\begin{figure}[htbp]
	\centering
	\setlength{\abovecaptionskip}{0.cm}
	\includegraphics[width=7cm]{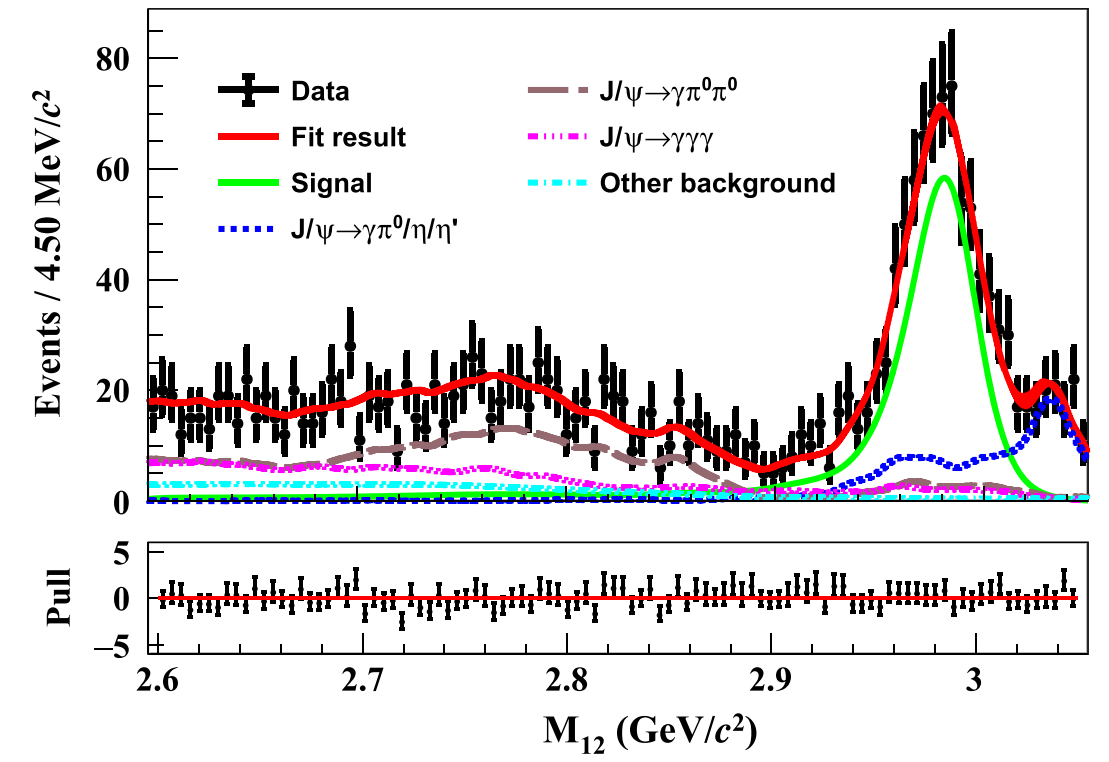}
	\caption{The invariant mass spectrum of highest energy combination of $\gamma\gamma$.}
	\label{fig:mass_etac_gamgam}
\end{figure}

Using $(2712.4\pm14.3)\times10^{6}$ $\psi(3686)$ events collected with the BESIII detector, the process $J/\psi\to\gamma\eta_c, \eta_c \to \gamma \gamma$ is observed via $\psi(3686)\to \pi^+\pi^- J/\psi$ for the first time~\cite{PhysRevLett.134.181901}. Compared to the directly produced $J/\psi$ sample which suffers from overwhelmingly high $e^+e^- \to \gamma_{\rm ISR}\gamma\gamma$ background, background level is significantly lower for this sample. After further suppression of backgrounds $J/\psi\to \gamma P(P=\pi^0,\eta,\eta^\prime)$ and $J/\psi\to\gamma\pi^0\pi^0$, clear $\eta_c$ signal is observed in the $M_{\gamma\gamma}$ spectrum as shown in Fig.~\ref{fig:mass_etac_gamgam}. By performing a one-dimensional fit, the $\eta_c$ signal yield is extracted to be $N_{\rm sig}=(677.7\pm33.5)$, and the product branching fraction $\mathcal{B}(J/\psi\to\gamma\eta_c)\times\mathcal{B}(\eta_c \to \gamma \gamma)$ is determined to be $(5.23\pm0.26_{\rm stat.}\pm0.30_{\rm syst.})\times10^{-6}$. Further comparison suggests good consistency between this measurement and most recent Lattice-QCD calculations~\cite{Colquhoun:2023zbc,Meng:2024axn} on the product branching fraction $\mathcal{B}(J/\psi\to\gamma\eta_c)\times\mathcal{B}(\eta_c \to \gamma \gamma)$. However, if one quote the PDG global fitted $\mathcal{B}(J/\psi\to\gamma\eta_c)$~\cite{ParticleDataGroup:2024cfk}, the resultant $\mathcal{B}(\eta_c \to \gamma \gamma)$ shows significant tension with both these theoretical works as well as PDG global fitted $\mathcal{B}(\eta_c \to \gamma \gamma)$.

\begin{figure}[htbp]
	\centering
	\setlength{\abovecaptionskip}{0.cm}
	\includegraphics[width=15cm]{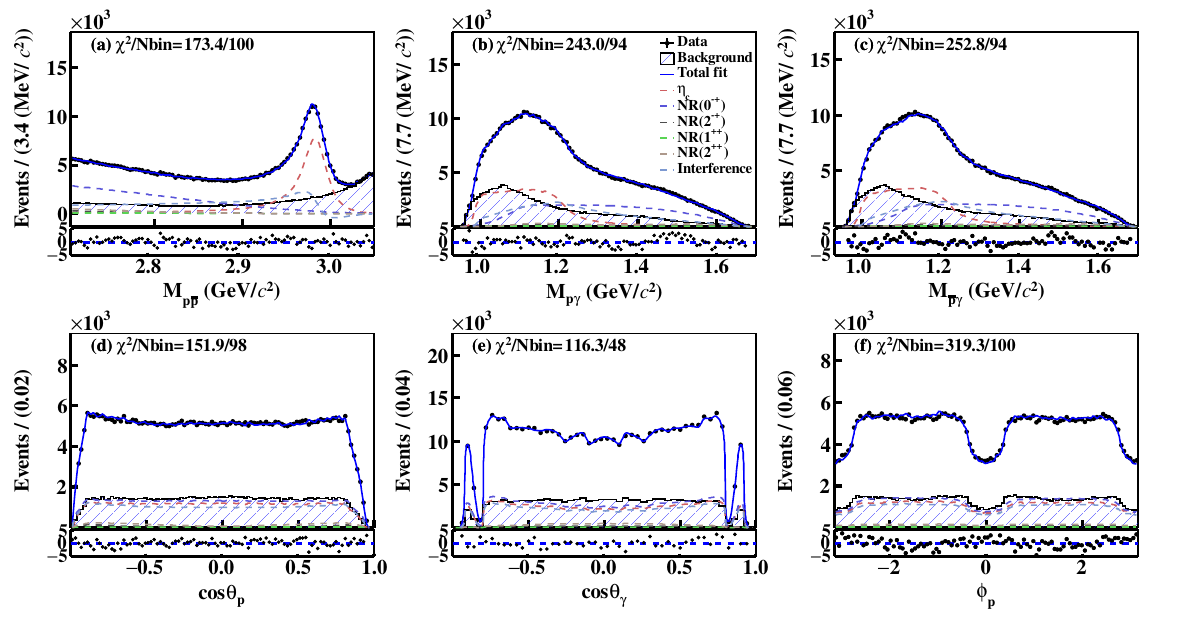}
	\caption{The amplitude fit results to $J/\psi\to\gamma p\bar{p}$ in the $\eta_c$ mass region.}
	\label{fig:amplitude_gampp}
\end{figure}

The PDG global fitted $\mathcal{B}(J/\psi\to\gamma\eta_c)$~\cite{ParticleDataGroup:2024cfk} are based on several measurements of $J/\psi\to\gamma\eta_c$ with $\eta_c$ decaying into various hadronic final states. Given the limited statistics in the previous measurements, the interference effects between $\eta_c$ and non-resonant components are generally ignored, which may result in some bias in the extracted yield hence the product branching fractions. This idea motivates a precision measurement of $J/\psi\to\gamma\eta_c,\eta_c \to p\bar{p}$~\cite{BESIII:2025vdn} with interference effects well considered. Using $(10.087\pm0.044)\times10^{9}$ $J/\psi$ events collected with the BESIII detector, the process $J/\psi\to\gamma p\bar{p}$ is selected with the main background contributions from $J/\psi\to\gamma^{\rm FSR} p\bar{p}$ and $p\bar{p}\pi^0$. Instead of one dimension fit, amplitude analysis is performed to extract the $\eta_c$ signal yields as shown in Fig.~\ref{fig:amplitude_gampp}, which provides more reliable interference pattern by distinguishing non-resonant components with difference spin-parities, and solve the multi-solution issue observed in one dimension fits. The resultant product branching fraction is determined to be $\mathcal{B}(J/\psi\to\gamma\eta_c)\times\mathcal{B}(\eta_c\to p\bar{p})=(2.11\pm0.02\pm0.07)\times10^{-5}$.

Further combination of three product branching fractions allows the determination of individual branching fractions as
\begin{equation}
\small
\begin{split}
\mathcal{B}(J/\psi\to\gamma\eta_{c})&=\sqrt{\frac{\left[\mathcal{B}(J/\psi\to\gamma\eta_c)\times\mathcal{B}(\eta_c\to p\bar{p})\right]\times \left[\mathcal{B}(J/\psi\to\gamma\eta_c)\times\mathcal{B}(\eta_c\to \gamma\gamma)\right]}{\left[\mathcal{B}(\eta_c\to p\bar{p})\times\mathcal{B}(\eta_c\to \gamma\gamma)\right]}}=(2.29\pm0.19)\%,\\
\mathcal{B}(\eta_{c}\to\gamma\gamma)&=\sqrt{\frac{\left[\mathcal{B}(\eta_c\to p\bar{p})\times\mathcal{B}(\eta_c\to \gamma\gamma)\right]\times\left[\mathcal{B}(J/\psi\to\gamma\eta_c)\times\mathcal{B}(\eta_c\to \gamma\gamma)\right]}{\left[\mathcal{B}(J/\psi\to\gamma\eta_c)\times\mathcal{B}(\eta_c\to p\bar{p})\right]}}=(2.28\pm0.19)\times10^{-4},\\
\mathcal{B}(\eta_{c}\to p\bar{p})&=\sqrt{\frac{\left[\mathcal{B}(\eta_c\to p\bar{p})\times\mathcal{B}(\eta_c\to \gamma\gamma)\right]\times\left[\mathcal{B}(J/\psi\to\gamma\eta_c)\times\mathcal{B}(\eta_c\to p\bar{p})\right]}{\left[\mathcal{B}(J/\psi\to\gamma\eta_c)\times\mathcal{B}(\eta_c\to \gamma\gamma)\right]}}=(0.92\pm0.08)\times10^{-3}.\\
\end{split}
\end{equation}
The obtained results shows good agreement with theoretical calculations~\cite{Colquhoun:2023zbc,Meng:2024axn} as shown in Fig.~\ref{fig:compare_th_exp_bfpdg}, which provide a promising solution to the first puzzle mentioned in the introduction.

\begin{figure}[htbp]
	\centering
	\setlength{\abovecaptionskip}{0.cm}
	\includegraphics[width=7cm]{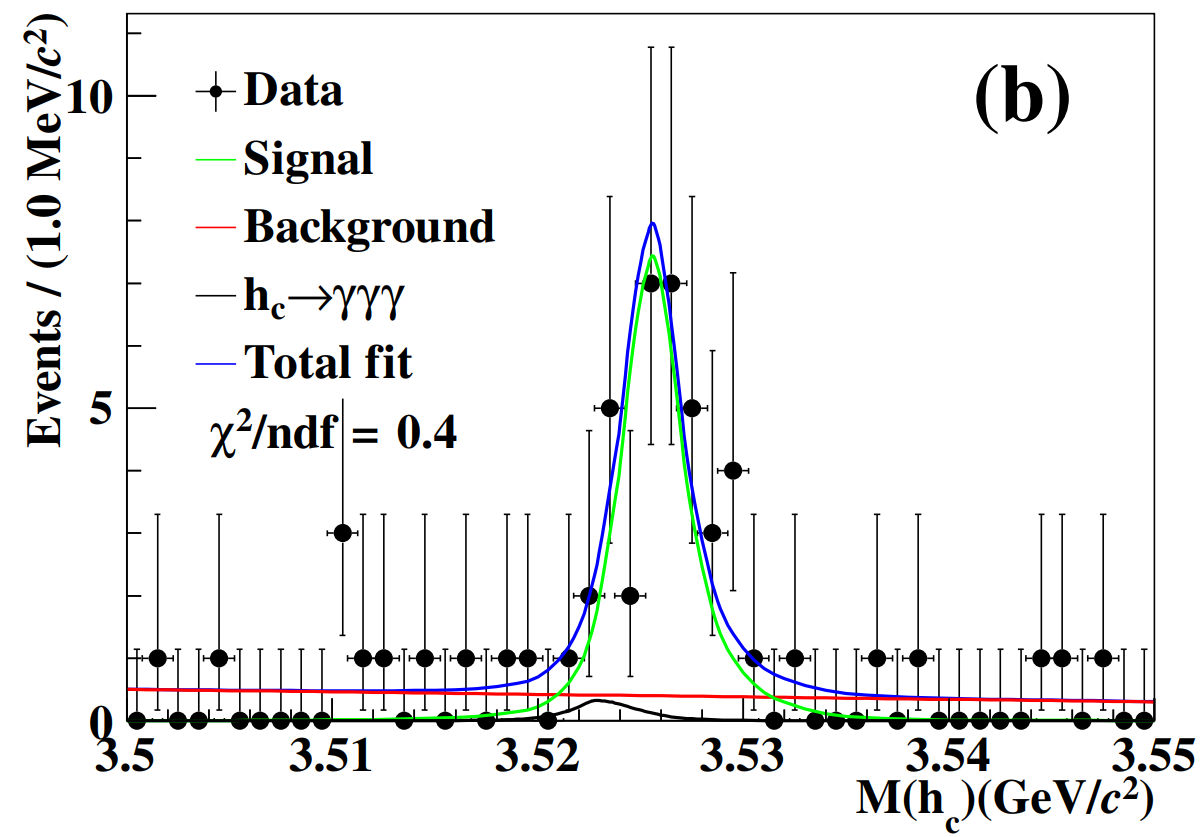}
	\caption{The invariant mass spectrum of $h_c$ with $h_c\to\gamma\eta_c,\eta_c \to \gamma\gamma$.}
	\label{fig:mass_hc_gametac}
\end{figure}

Further measurement of $\eta_c\to\gamma\gamma$ is performed in $\psi(3686)\to\pi^0 h_c, h_c\to\gamma\eta_c$~\cite{BESIII:2026pff}. Benefiting from the low background level, absolute branching fraction measurement is performed by tagging $\eta_c$ in the recoiling system of $(\pi^0\gamma)_{\rm tag}$ with $N_{\rm tag}\sim 1.6\times 10^5$. Based on the selected $(\pi^0\gamma)_{\rm tag}$, signal $\eta_c \to \gamma\gamma$ is reconstructed, with fitted signal yield $N_{\rm sig}=(32.5\pm6.4)$ as shown in Fig.~\ref{fig:mass_hc_gametac}. The resultant branching fraction is determined to be $\mathcal{B}(\eta_c \to \gamma\gamma)=(2.45\pm0.48_{\rm stat.}\pm0.09_{\rm syst.})\times10^{-4}$, which is consistent with both combined results~\cite{BESIII:2025vdn} above as well as the theoretical works~\cite{Colquhoun:2023zbc,Meng:2024axn}. The similar approach could also be applied to measure the absolute branching fractions of hadronic decay mode of $\eta_c$.

\section{Measurement of $\eta_c$ decaying into hadronic modes}

To reveal the missed half decay modes of $\eta_c$, several recent first/updated measurements and searches of $\eta_c$ hadronic modes are performed at BESIII, as listed here:

\begin{itemize}
	\item Using $(2712.4\pm14.3)\times10^{6}$ $\psi(3686)$ events collected with the BESIII detector, the branching fraction $\mathcal{B}(\eta_c\to 2(\pi^+\pi^-)\eta)$~\cite{BESIII:2024rdn} is measured to be $(5.5\pm0.5\pm1.9)\%$ and $(2.6\pm0.4\pm1.3)\%$ with improved statistics in $\psi(3686)\to\gamma\eta_c$. Here, the two in-distinguishable solutions correspond to destructive and constructive interference respectively.
	
	\item Using $(10.087\pm0.044)\times10^{9}$ $J/\psi$ events collected with the BESIII detector, the branching fraction $\mathcal{B}(\eta_c\to \Xi^0 \bar{\Xi}^0)$~\cite{BESIII:2026nsk} is measured to be $(1.63\pm0.22)\%$ and $(1.33\pm0.20)\%$ for the first time in $J/\psi\to\gamma\eta_c$.
	\item Using $(2712.4\pm14.3)\times10^{6}$ $\psi(3686)$ events collected with the BESIII detector, the isospin violated process $\eta_c\to \Lambda\bar{\Sigma}^0+c.c.$~\cite{BESIII:2026jrd} is searched for the first time, no significant signal is seen and upper limit is set to be $\mathcal{B}(\eta_c\to \Lambda\bar{\Sigma}^0+c.c.)<6.2\times10^{-5}$.
\end{itemize}

\section{Summary}
  
Using 10 Billion $J/\psi$ and 2.7 Billion $\psi(3686)$ data sets collected by the BESIII detector, we present precision measurements on the branching fractions $\mathcal{B}(J/\psi\to\gamma\eta_c)$ and $\mathcal{B}(\eta_c\to\gamma\gamma)$ with uncertainties less than 10\%. The obtained results are in good agreement with most recent Lattice QCD calculations, hence provide a promising solution for the discrepancy between theoretical and experimental works. The absolute branching fraction $\mathcal{B}(\eta_c\to\gamma\gamma)$ is also measured in $\psi(3686)\to\pi^0 h_c, h_c \to \gamma\eta_c$, and the obtained results is consistent with $J/\psi\to\gamma\eta_c$ production. Several hadronic decay modes $\eta_c\to2(\pi^+\pi^-)\eta$, $\eta_c\to \Xi^0\bar{\Xi}^0$, and $\eta_c \to \Lambda\bar{\Sigma}^0+c.c.$ are also measured.

\section{Acknowledgments}

This work is supported by the National Natural Science Foundation of China~(NSFC) under Contracts No. 12192264.

\section*{References}
\bibliography{moriond}

@article{E598:1974sol,
    author = "Aubert, J. J. and others",
    collaboration = "E598",
    title = "{Experimental Observation of a Heavy Particle $J$}",
    reportNumber = "COO-3069-271",
    doi = "10.1103/PhysRevLett.33.1404",
    journal = "Phys. Rev. Lett.",
    volume = "33",
    pages = "1404--1406",
    year = "1974"
}

@article{SLAC-SP-017:1974ind,
    author = "Augustin, J. E. and others",
    collaboration = "SLAC-SP-017",
    title = "{Discovery of a Narrow Resonance in $e^+ e^-$ Annihilation}",
    reportNumber = "SLAC-PUB-1504, LBL-3391",
    doi = "10.1103/PhysRevLett.33.1406",
    journal = "Phys. Rev. Lett.",
    volume = "33",
    pages = "1406--1408",
    year = "1974"
}

@article{ParticleDataGroup:2024cfk,
    author = "Navas, S. and others",
    collaboration = "Particle Data Group",
    title = "{Review of particle physics}",
    doi = "10.1103/PhysRevD.110.030001",
    journal = "Phys. Rev. D",
    volume = "110",
    number = "3",
    pages = "030001",
    year = "2024"
}

@article{Godfrey:1985xj,
    author = "Godfrey, S. and Isgur, Nathan",
    title = "{Mesons in a Relativized Quark Model with Chromodynamics}",
    doi = "10.1103/PhysRevD.32.189",
    journal = "Phys. Rev. D",
    volume = "32",
    pages = "189--231",
    year = "1985"
}

@article{Dudek:2006ej,
    author = "Dudek, Jozef J. and Edwards, Robert G. and Richards, David G.",
    title = "{Radiative transitions in charmonium from lattice QCD}",
    eprint = "hep-ph/0601137",
    archivePrefix = "arXiv",
    reportNumber = "JLAB-THY-06-457",
    doi = "10.1103/PhysRevD.73.074507",
    journal = "Phys. Rev. D",
    volume = "73",
    pages = "074507",
    year = "2006"
}

@article{Dudek:2009kk,
    author = "Dudek, Jozef J. and Edwards, Robert and Thomas, Christopher E.",
    title = "{Exotic and excited-state radiative transitions in charmonium from lattice QCD}",
    eprint = "0902.2241",
    archivePrefix = "arXiv",
    primaryClass = "hep-ph",
    reportNumber = "JLAB-THY-09-949",
    doi = "10.1103/PhysRevD.79.094504",
    journal = "Phys. Rev. D",
    volume = "79",
    pages = "094504",
    year = "2009"
}

@article{Chen:2011kpa,
    author = "Chen, Ying and others",
    title = "{Radiative transitions in charmonium from $N_f=2$ twisted mass lattice QCD}",
    eprint = "1104.2655",
    archivePrefix = "arXiv",
    primaryClass = "hep-lat",
    doi = "10.1103/PhysRevD.84.034503",
    journal = "Phys. Rev. D",
    volume = "84",
    pages = "034503",
    year = "2011"
}

@article{Becirevic:2012dc,
    author = "Becirevic, Damir and Sanfilippo, Francesco",
    title = "{Lattice QCD study of the radiative decays $J/\psi\to \eta_c\gamma$ and $h_c\to \eta_c\gamma$}",
    eprint = "1206.1445",
    archivePrefix = "arXiv",
    primaryClass = "hep-lat",
    reportNumber = "LPT-12-52",
    doi = "10.1007/JHEP01(2013)028",
    journal = "JHEP",
    volume = "01",
    pages = "028",
    year = "2013"
}

@article{Donald:2012ga,
    author = "Donald, G. C. and Davies, C. T. H. and Dowdall, R. J. and Follana, E. and Hornbostel, K. and Koponen, J. and Lepage, G. P. and McNeile, C.",
    title = "{Precision tests of the $J/{\psi}$ from full lattice QCD: mass, leptonic width and radiative decay rate to ${\eta}_c$}",
    eprint = "1208.2855",
    archivePrefix = "arXiv",
    primaryClass = "hep-lat",
    doi = "10.1103/PhysRevD.86.094501",
    journal = "Phys. Rev. D",
    volume = "86",
    pages = "094501",
    year = "2012"
}

@article{Gui:2019dtm,
    author = "Gui, Long-Cheng and Dong, Jia-Mei and Chen, Ying and Yang, Yi-Bo",
    title = "{Study of the pseudoscalar glueball in $J/\psi$ radiative decays}",
    eprint = "1906.03666",
    archivePrefix = "arXiv",
    primaryClass = "hep-lat",
    doi = "10.1103/PhysRevD.100.054511",
    journal = "Phys. Rev. D",
    volume = "100",
    number = "5",
    pages = "054511",
    year = "2019"
}

@article{Colquhoun:2023zbc,
    author = "Colquhoun, Brian and Cooper, Laurence J. and Davies, Christine T. H. and Lepage, G. Peter",
    collaboration = "Particle Data Group, HPQCD, (HPQCD Collaboration){\textdaggerdbl}",
    title = "{Precise determination of decay rates for $\eta_c\to\gamma \gamma$, $J/\psi\to\gamma\eta_c$, $J/\psi\to e^+e^-\eta_c$ from lattice QCD}",
    eprint = "2305.06231",
    archivePrefix = "arXiv",
    primaryClass = "hep-lat",
    doi = "10.1103/PhysRevD.108.014513",
    journal = "Phys. Rev. D",
    volume = "108",
    number = "1",
    pages = "014513",
    year = "2023"
}

@article{Meng:2024axn,
    author = "Meng, Yu and Liu, Chuan and Wang, Teng and Yan, Haobo",
    title = "{Lattice study of $J/\psi\to\gamma\eta_c$ using a method without momentum extrapolation}",
    eprint = "2411.04415",
    archivePrefix = "arXiv",
    primaryClass = "hep-lat",
    doi = "10.1103/PhysRevD.111.014508",
    journal = "Phys. Rev. D",
    volume = "111",
    number = "1",
    pages = "014508",
    year = "2025"
}

@article{Brambilla:2005zw,
    author = "Brambilla, Nora and Jia, Yu and Vairo, Antonio",
    title = "{Model-independent study of magnetic dipole transitions in quarkonium}",
    eprint = "hep-ph/0512369",
    archivePrefix = "arXiv",
    reportNumber = "IFUM-841-FT",
    doi = "10.1103/PhysRevD.73.054005",
    journal = "Phys. Rev. D",
    volume = "73",
    pages = "054005",
    year = "2006"
}

@article{Pineda:2013lta,
    author = "Pineda, Antonio and Segovia, J.",
    title = "{Improved determination of heavy quarkonium magnetic dipole transitions in potential nonrelativistic QCD}",
    eprint = "1302.3528",
    archivePrefix = "arXiv",
    primaryClass = "hep-ph",
    doi = "10.1103/PhysRevD.87.074024",
    journal = "Phys. Rev. D",
    volume = "87",
    number = "7",
    pages = "074024",
    year = "2013"
}

@article{Segovia:2021bjb,
    author = "Segovia, Jorge and Tarr{\'u}s Castell{\`a}, Jaume",
    title = "{Line shape and the experimental determination of the $J/\psi \to \gamma\eta_c$ branching fraction}",
    eprint = "2106.15203",
    archivePrefix = "arXiv",
    primaryClass = "hep-ph",
    doi = "10.1103/PhysRevD.104.074032",
    journal = "Phys. Rev. D",
    volume = "104",
    number = "7",
    pages = "074032",
    year = "2021"
}

@article{Beilin:1984pf,
    author = "Beilin, V. A. and Radyushkin, A. V.",
    title = "{Quantum Chromodynamic Sum Rules and $J/\psi \to \eta_c \gamma$ Decay}",
    reportNumber = "JINR-P2-84-557",
    doi = "10.1016/0550-3213(85)90310-4",
    journal = "Nucl. Phys. B",
    volume = "260",
    pages = "61--78",
    year = "1985"
}

@article{Guo:2019xqa,
    author = "Guo, Song-Pei and Sun, Yan-Jun and Hong, Wei and Huang, Qi and Zhao, Guo-Hua",
    title = "{Study of the radiative decay $J/\psi\to\gamma\eta_c$ in light cone sum rules}",
    eprint = "1912.00836",
    archivePrefix = "arXiv",
    primaryClass = "hep-ph",
    doi = "10.1016/j.nuclphysb.2020.115053",
    journal = "Nucl. Phys. B",
    volume = "955",
    pages = "115053",
    year = "2020"
}

@article{Khodjamirian:1979fa,
    author = "Khodjamirian, A. Yu.",
    title = "{Dispersion Sum Rules for the Amplitudes of Radiative Transitions in Quarkonium}",
    reportNumber = "EFI-376-34-79-YEREVAN",
    doi = "10.1016/0370-2693(80)90974-0",
    journal = "Phys. Lett. B",
    volume = "90",
    pages = "460--464",
    year = "1980"
}

@article{Ebert:2002pp,
    author = "Ebert, D. and Faustov, R. N. and Galkin, V. O.",
    title = "{Properties of heavy quarkonia and $B_c$ mesons in the relativistic quark model}",
    eprint = "hep-ph/0210381",
    archivePrefix = "arXiv",
    reportNumber = "HU-EP-02-45",
    doi = "10.1103/PhysRevD.67.014027",
    journal = "Phys. Rev. D",
    volume = "67",
    pages = "014027",
    year = "2003"
}

@article{Deng:2016stx,
    author = "Deng, Wei-Jun and Liu, Hui and Gui, Long-Cheng and Zhong, Xian-Hui",
    title = "{Charmonium spectrum and their electromagnetic transitions with higher multipole contributions}",
    eprint = "1608.00287",
    archivePrefix = "arXiv",
    primaryClass = "hep-ph",
    doi = "10.1103/PhysRevD.95.034026",
    journal = "Phys. Rev. D",
    volume = "95",
    number = "3",
    pages = "034026",
    year = "2017"
}

@article{Barnes:2005pb,
    author = "Barnes, T. and Godfrey, S. and Swanson, E. S.",
    title = "{Higher charmonia}",
    eprint = "hep-ph/0505002",
    archivePrefix = "arXiv",
    doi = "10.1103/PhysRevD.72.054026",
    journal = "Phys. Rev. D",
    volume = "72",
    pages = "054026",
    year = "2005"
}

@article{BESIII:2021cxx,
    author = "Ablikim, M. and others",
    collaboration = "BESIII",
    title = "{Number of $J/\psi$ events at BESIII}",
    eprint = "2111.07571",
    archivePrefix = "arXiv",
    primaryClass = "hep-ex",
    doi = "10.1088/1674-1137/ac5c2e",
    journal = "Chin. Phys. C",
    volume = "46",
    number = "7",
    pages = "074001",
    year = "2022"
}

@article{BESIII:2024lks,
    author = "Ablikim, Medina and others",
    collaboration = "BESIII",
    title = "{Determination of the number of $\psi(3686)$ events taken at BESIII*}",
    eprint = "2403.06766",
    archivePrefix = "arXiv",
    primaryClass = "hep-ex",
    doi = "10.1088/1674-1137/ad595b",
    journal = "Chin. Phys. C",
    volume = "48",
    number = "9",
    pages = "093001",
    year = "2024"
}

@article{PhysRevLett.134.181901,
  title = {Observation of the Charmonium Decay $\eta_c \to \gamma\gamma$ in $J/\psi\to\gamma\eta_c$},
  author = "Ablikim, Medina and others",
  collaboration = {BESIII Collaboration},
  journal = {Phys. Rev. Lett.},
  volume = {134},
  issue = {18},
  pages = {181901},
  numpages = {10},
  year = {2025},
  month = {May},
  publisher = {American Physical Society},
  doi = {10.1103/PhysRevLett.134.181901},
  url = {https://link.aps.org/doi/10.1103/PhysRevLett.134.181901}
}

@article{BESIII:2025vdn,
    author = "Ablikim, Medina and others",
    collaboration = "BESIII",
    title = "{Study of the Magnetic Dipole Transition of $J/\psi\to\gamma\eta_c$ via $\eta_c \to p\bar{p}$}",
    eprint = "2510.15247",
    archivePrefix = "arXiv",
    primaryClass = "hep-ex",
    doi = "10.1103/r6yf-q6kv",
    journal = "Phys. Rev. Lett.",
    volume = "136",
    number = "5",
    pages = "051901",
    year = "2026"
}

@article{BESIII:2026pff,
    author = "Ablikim, Medina and others",
    collaboration = "BESIII",
    title = "{First Measurement of the Absolute Branching Fraction of $\eta_c \to \gamma \gamma$}",
    journal = "arXiv 2601.11236"
}

@article{BESIII:2024rdn,
    author = "Ablikim, Medina and others",
    collaboration = "BESIII",
    title = "{Observation of $\eta_c \to 2(\pi^+\pi^-)\eta$ via $\psi(3686)$ radiative transitions}",
    eprint = "2406.08225",
    archivePrefix = "arXiv",
    primaryClass = "hep-ex",
    doi = "10.1103/PhysRevD.111.052013",
    journal = "Phys. Rev. D",
    volume = "111",
    number = "5",
    pages = "052013",
    year = "2025"
}

@article{BESIII:2026nsk,
    author = "Ablikim, Medina and others",
    collaboration = "BESIII",
    title = "{First observation of the $\eta_c \to \Xi^0 \bar{\Xi}^0$ decay}",
    eprint = "2602.09652",
    archivePrefix = "arXiv",
    primaryClass = "hep-ex",
    doi = "10.1103/ckyx-hyjm",
    journal = "Phys. Rev. D",
    volume = "113",
    number = "7",
    pages = "072003",
    year = "2026"
}

@article{BESIII:2026jrd,
    author = "Ablikim, Medina and others",
    collaboration = "BESIII",
    title = "{Search for the isospin-violating decays $\chi_{cJ}\to \Lambda\bar{\Sigma}^0+c.c.$ and $\eta_c\to \Lambda\bar{\Sigma}^0+c.c.$}",
    journal = "arXiv 2601.19493"
}

@article{CLQCD:2020njc,
    author = "Chen, Ying and Gong, Ming and Li, Ning and Liu, Chuan and Liu, Yu-Bin and Liu, Zhaofeng and Ma, Jian-Ping and Meng, Yu and Xiong, Chao and Zhang, Ke-Long",
    collaboration = "CLQCD",
    title = "{Lattice study of two-photon decay widths for scalar and pseudo-scalar charmonium}",
    eprint = "2003.09817",
    archivePrefix = "arXiv",
    primaryClass = "hep-lat",
    doi = "10.1088/1674-1137/44/8/083108",
    journal = "Chin. Phys. C",
    volume = "44",
    number = "8",
    pages = "083108",
    year = "2020",
    note = "[Erratum: Chin.Phys.C 46, 059001 (2022)]"
}

@article{CLQCD:2016ugl,
    author = "Chen, Ting and others",
    collaboration = "CLQCD",
    title = "{Two-photon decays of $\eta _c$ from lattice QCD}",
    eprint = "1602.00076",
    archivePrefix = "arXiv",
    primaryClass = "hep-lat",
    doi = "10.1140/epjc/s10052-016-4212-8",
    journal = "Eur. Phys. J. C",
    volume = "76",
    number = "7",
    pages = "358",
    year = "2016"
}

@article{Liu:2020qfz,
    author = "Liu, Chuan and Meng, Yu and Zhang, Ke-Long",
    title = "{Ward identity of the vector current and the decay rate of $\eta_c\rightarrow\gamma\gamma$ in lattice QCD}",
    eprint = "2004.03907",
    archivePrefix = "arXiv",
    primaryClass = "hep-lat",
    doi = "10.1103/PhysRevD.102.034502",
    journal = "Phys. Rev. D",
    volume = "102",
    number = "3",
    pages = "034502",
    year = "2020"
}
\end{document}